\documentclass{emulateapj}
\providecommand{\mb}{$\Delta m_{15}(B)$}

\providecommand{\mu}{$\Delta m_{15}(U)$}

\begin{document}
%%%%%%%%%% Title & Author %%%%%%%%%%%%%%%%%%%%%%%%%%%%%%%%%%%%%%%%%%%%%%%%%%%%%%%
\title{Theoretical Clues on the Ultraviolet Diversity of Type I\lowercase{a} Supernovae}
%\author{Peter~J.~Brown}
%\affil{George P. and Cynthia Woods Mitchell Institute for Fundamental Physics \& Astronomy, \\
%Texas A. \& M. University, Department of Physics and Astronomy, \\
%4242 TAMU, College Station, TX 77843, USA }            

%\email{pbrown@physics.tamu.edu}  

\author{Peter~J.~Brown\altaffilmark{1}, 
E. Baron\altaffilmark{2}, 
%Stephen Holland\altaffilmark{3,4}, 
Peter Milne\altaffilmark{3}, \\ 
Peter W. A. Roming\altaffilmark{4,5}, 
%Emma Walker\altaffilmark{8,9}, 
\& Lifan Wang\altaffilmark{1}
}
%\altaffiltext{1}{Department of Physics \& Astronomy, University of Utah, 115 South 1400 East \#201, 
%Salt Lake City, UT 84112, USA}            
\altaffiltext{1}{George P. and Cynthia Woods Mitchell Institute for Fundamental Physics \& Astronomy, 
Texas A. \& M. University, Department of Physics and Astronomy, 
4242 TAMU, College Station, TX 77843, USA; 
pbrown@physics.tamu.edu}            

\altaffiltext{2}{Homer L. Dodge Department of Physics and Astronomy, University of Oklahoma, 440 W. Brooks, Rm 100, Norman, OK 73019-2061, USA}

%\altaffiltext{3}{Space Telescope Science Center
%3700 San Martin Dr., 
%Baltimore, MD 21218, USA}
%\altaffiltext{4}{Astrophysics Science Division, Code 660.1,
%                8800 Greenbelt Road
%                Goddard Space Flight Centre,
%                Greenbelt, MD 20771, USA}
\altaffiltext{3}{Steward Observatory, University of Arizona, Tucson, AZ 85719, USA}            
\altaffiltext{4}{ Southwest Research Institute, 6220 Culebra Road, San Antonio, TX 78238, USA}            
\altaffiltext{5}{ University of Texas at San Antonio, 1 UTSA Circle, San Antonio, TX, 78249}            
%\altaffiltext{8}{Yale University, New Haven, CT, USA}            
%\altaffiltext{9}{Scuola Normale Superiore, 7, I-56126 Pisa, Italy }            

%%%%%%%%%% Abstract %%%%%%%%%%%%%%%%%%%%%%%%%%%%%%%%%%%%%%%%%%%%%%%%%%%%%%%%%%%%
\begin{abstract}

The effect of metallicity on the observed light of Type Ia supernovae (SNe Ia) 
could lead to systematic errors as the absolute magnitudes of local and distant SNe Ia 
are compared to measure luminosity distances and determine cosmological parameters.  
The UV light may be especially sensitive to metallicity, though different modeling 
methods disagree as to the magnitude, wavelength dependence, and even the sign of the effect.  
The outer density structure, and to a lesser degree asphericiy, also impact the UV.  We compute synthetic photometry of various metallicity-dependent models and 
compare to UV/optical photometry from the Swift Ultra-Violet/Optical Telescope.  
We find that the scatter in the mid-UV to near-UV colors is larger than predicted 
by changes in metallicity alone and is not consistent with reddening.  
We demonstrate that a recently employed method to determine relative abundances using UV spectra 
can be done using UVOT photometry, but we warn that accurate results require an accurate model of the cause of the variations. The abundance of UV photometry 
now available should provide constraints on models that typically rely on UV spectroscopy 
for constraining metallicity, density, and other parameters.  Nevertheless, UV spectroscopy for a variety of SN explosions is still needed to guide the creation of accurate models.  A better understanding 
of the influences affecting the UV is important for using SNe Ia as cosmological probes, 
as the UV light may test whether SNe Ia are significantly affected by evolutionary effects.

\end{abstract}

%%%%%%%%%% Keywords %%%%%%%%%%%%%%%%%%%%%%%%%%%%%%%%%%%%%%%%%%%%%%%%%%%%%%%%%%%%
%%%%%%%%%%%%%%%%%%%%%%%%%%%%%%%%%%%%%%%%%%%%%%%%%%%%%%%%%%%%%%%%%%%%%%%%%%%%%%%%%
%\clearpage

\section{The Influence of Metallicity on Type I\lowercase{a} Supernovae  \label{intro}}
 
Supernovae are important cosmological tools for measuring the expansion history of the universe 
\citep{Riess_etal_1998,Schmidt_etal_1998,Perlmutter_etal_1999, Suzuki_etal_2012, Ganeshalingam_etal_2013}.  Their usefulness as standardizable candles (c.f. \citealp{Branch_1998,Leibundgut_2001}) requires the relationships between the luminosity and the observed color and light curve shape to be the same for SNe at all redshifts.
One major concern is that evolution in the properties of the progenitors with redshift will systematically change their luminosity and lead to incorrect distance measurements. One property expected to change with redshift is the average metal content (mass fraction of elements heavier than helium) of the universe and its constituents as stars form from the elements created in the lives and deaths of previous generations of stars.  We will refer to the metal content generically as `metallicity' but will clarify as appropriate the specific elemental ratios or abundances.

Metallicity could affect the progenitor system and explosion (including the SN observables) at all times leading up to, during, and after the explosion. The premordial metallicity at the time of the SN progenitor formation could affect the evolution of the SN progenitor into a white dwarf 
and its final C/O ratio and central density \citep{Umeda_etal_1999}.  It could affect the mass lost by the accreting white dwarf through winds \citep{Kobayashi_etal_1998}.  The metallicity of the donated material could change the metallicity of the outer layers of the white dwarf.  The metallicity of the WD could affect its final density structure.  The electron fraction (influenced by the amount of $^{22}$Ne present in the progenitor) affects the relative abundances of $^{56}$Ni (which powers the luminosity), $^{58}$Ni, and $^{54}$Fe \citep{Mazzali_Podsiadlowski_2006,Hoeflich_etal_2013}.  
%%%%%%%%%% fix the Hoeflich reference
The final metallicity (and its spatial variation) within the ejecta could be affected not just by the primordial metallicity but by the amount of nucleosynthetic products of the explosion and the degree to which they are mixed.  

By its observational nature, astrophysicists cannot change the initial conditions of actual stars and 
observe how their explosions and observable characteristics respond.  Instead, theoretical models are made from the best available knowledge of conditions and physical properties.  
The ways that metallicity can be incorporated into theoretical models vary greatly.
\citet{Hoeflich_etal_1998} modified the pre-explosion metal content and examined the effect on light curves.
\citet{Lentz_etal_2000} attempted to replicate this in a radiation transport simulation by scaling the amount of elements heavier than oxygen in the unburned layers and the amount of $^{54}$Fe in the incomplete burning zone.
\citet{Timmes_etal_2003} studied the relationship between Z and the $^{56}$Ni mass (and thus the luminosity).  \citet{Sauer_etal_2008} tested the effect of varying the amount of stable Fe, $^{56}$Ni, and Ti and Cr together.
\citet{Bravo_etal_2010} changed the metallicity Z in the pre-main-sequence model, evolved it and exploded it to study the mass of $^{56}$Ni ejected, the corresponding bolometric luminosity, and the effect on the luminosity-width relation.
Most of these studies show that the effect of changing heavy metal abundances is the strongest at UV wavelengths \citep{Lentz_etal_2000,Sauer_etal_2008,Walker_etal_2012}.

Observationally, many studies have looked for `metallicity' or similar effects in the properties of SNe 
by comparing with properties of the host galaxy.  Most focus on two properties of SNe -- their peak luminosity (which is correlated with the light curve shape) and the difference between the peak luminosity and that expected based on the light curve shape.  The latter is often given in terms of Hubble residuals (HR), defined as the difference between the redshift distance modulus (calculated using the redshift of the host galaxy and the best fit cosmology) and the luminosity distance modulus (calculated from the SN flux and light curve shape). 
\citet{Gallagher_etal_2005} found no correlation between emission line metallicity of star forming host galaxies and the SN luminosity and a low significance correlation with HR.  They concluded that metallicity could be a secondary effect. 
\citet{Gallagher_etal_2008} reported a correlation between age and optical luminosity and between HR and the metal abundance for early type galaxy hosts using diagnostic grids of spectral indices to determine the age and metallicity.  \citet{Howell_etal_2009} were unable to reproduce the latter result using data from the Supernova Legacy Survey (SNLS).  They used SED fitting to detrmine the galaxy masses and the \citet{Tremonti_etal_2004} relation to convert to metallicity.   \citet{Howell_etal_2009} did find a correlation between the SN $^{56}$Ni mass (determined from the peak bolometric luminosity of the SN) and the luminosity-weighted age of the host.  For nearby SNe \citep{Kelly_etal_2010} as well as more distant SNe with SNLS data \citep{Sullivan_etal_2010} and SDSS-II data \citep{Lampeitl_etal_2010}, a relationship was found between the light curve shape-corrected SN luminosity and the mass of the host galaxy (estimated from SED fitting).  \citet{Dandrea_etal_2011} found significant differences when correlating HR with the gas phase metallicity or the specific star formation rate.  The underlying source of the correlations is unclear, as the host galaxy mass, metallicity and star formation are 
all connected together.  \citet{Hayden_etal_2013} compare the residual scatter after correcting for mass alone and after correcting using the fundamental metallicity relation (which incorporates mass and star formation).  They find the fundamental metallicity relation provides a significantly better correction and conclude that metallicity is the primary cause of the SN variations.  \citet{Childress_etal_2013_HR} find HR differences at low and high host mass with a sharp break in between.  They argue that dust and progenitor age could also explain the differences.

Despite the larger expected differences in the UV, tests at those wavelengths have lagged behind due to the relative paucity of UV SN data compared to that available in the optical.
Rest-frame near-UV spectra of local SNe were compared to more distant samples by \citet{Riess_etal_2007}, using rest-frame near-UV spectra of z$\sim$1.1 SNe observed with HST.  \citet{Ellis_etal_2008}, using z$\sim$0.5 SNe from SNLS, \citet{Foley_etal_2008} using z$\sim$0.5 SNe from ESSENCE, and \citet{Foley_etal_2012_U}, using z$\sim$0.2 SNe from SDSSII.  In constructing a mean spectrum, all noted an increase in the average flux level in the UV and an increase in the dispersion at shorter wavelengths.  \citet{Ellis_etal_2008} found the variations to be larger than that expected from metallicity differences as modeled by \citet{Lentz_etal_2000}.  \citet{Foley_etal_2008} noted that the blueshifted Si II 6355 \AA~would indicate higher metallicity \citep{Lentz_etal_2000}, the weaker Fe III 5129 \AA ~line could indicate lower metallicity \citep{Hoeflich_etal_1998,Sauer_etal_2008}, and the higher UV flux level could indicate higher \citep{Hoeflich_etal_1998} or lower metallicity \citep{Lentz_etal_2000,Sauer_etal_2008}.  
UV photometric studies 
%with Swift's Ultraviolet/Optical Telescope (UVOT; \citealp{Gehrels_etal_2004,Roming_etal_2005}) 
showed a modest increase in scatter at near-UV wavelengths \citep{Brown_etal_2010,Wang_etal_2012,Milne_etal_2013} and a large dispersion in the mid-UV \citep{Brown_etal_2010}.  Recently, \citet{Foley_Kirshner_2013} presented a relative metallicity determination between two SNe Ia 
based on UV spectra from HST which was further discussed in \citet{Graham_etal_2015}.

In examining the possibility of evolution in SNe Ia, \citet{Howell_etal_2007} and \citet{Sullivan_etal_2009} find modest evolution in the optical properties which may be of concern when much higher accuracies are demanded to differentiate dark energy models.  \citet{Sullivan_etal_2009} also find a UV flux difference between the low-z and intermediate-z SN samples, but the nearby sample only had 3 spectra within the SN stretch and epoch cuts.  
\citet{Cooke_etal_2011} and \citet{Maguire_etal_2012} used a much larger sample of near-UV spectra from HST, confirming the earlier results.  Additionally, \citet{Maguire_etal_2012} used the models from \citet{Walker_etal_2012} to conclude that the magnitude of both the dispersion and evolution were consistent with variation and evolution in the metallicity.
\citet{Milne_etal_2015} categorize local and distant SNe Ia based on the NUV-blue/red dichotomy demonstrated in \citet{Milne_etal_2013}.  They find that the systematic differences in the UV flux of nearby and distant type Ia SNe discussed above results not from a shift in the color distribution but a shift in the relative fractions of type Ia belonging to these two groups whose color distributions are similar across redshifts.

%As the UV is more strongly affected by metallicity than the optical, it could serve as a ``canary in a coal mine'' to warn about potential biases before they become significant enough in the optical.

%In Section \ref{obs} we will discuss the observational sample.  The theoretical models to be utilized will be described in Section \ref{models}.  In Section \ref{results} we will compare the observations to the models using colors and color differences.  Our conclusions are summarized and discussed in Section \ref{conclusion}.

With high quality UV spectra from HST and a large photometric sample from Swift/UVOT, it seems timely to access some of the SN models and how they relate to now available observations.
In Section 2 we will discuss the observational sample.  The theoretical models to be utilized will be described in Section 3.  In Section 4 we will compare the observations to the models using colors and color differences.  Our conclusions are summarized and discussed in Section 5.

\section{Swift Observations  \label{obs}} 
 
\begin{figure} 
\plotone{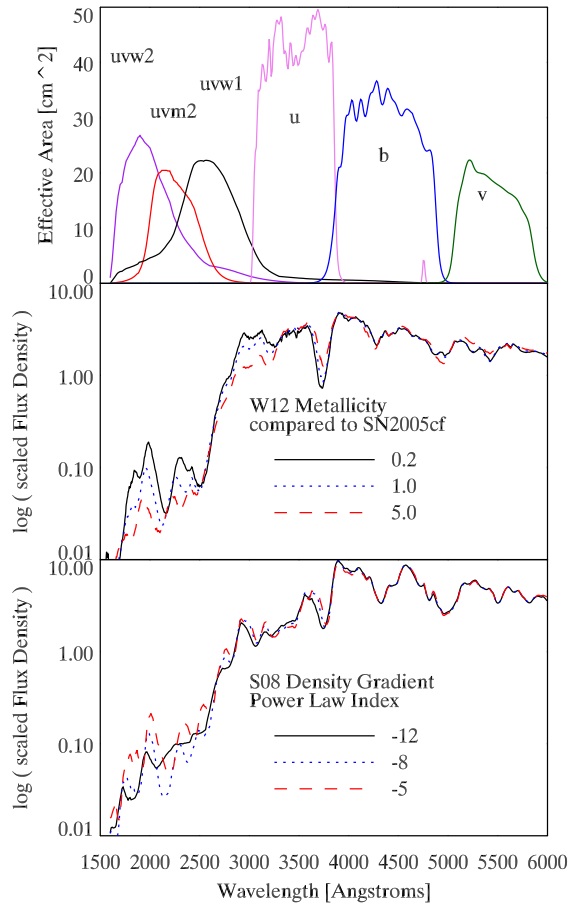} 
\caption[UVOT Filter Range]
        {Top Panel:  Swift UVOT filter curves.  
Middle Panel: Theoretical spectra from \citet{Walker_etal_2012} for which the metal content has been scaled relative to the best-fit model for SN~2005cf.  
Bottom Panel: Theoretical spectra from \citet{Sauer_etal_2008} for which the outer density gradient has been parameterized by a power-law and the index varied.
 } \label{fig_filters}    
\end{figure}

%\subsection{Data Reduction}
For our observational sample, we utilize observations of SNe Ia from the Swift Ultra-violet/Optical Telescope \citep{Gehrels_etal_2004,Roming_etal_2005}.  We use SNe previously published \citep{Brown_etal_2009,Milne_etal_2010,Brown_etal_2012_11fe,Brown_etal_2012_shock, Milne_etal_2013}, but updating the previously published photometry as necessary with the revised UV zeropoints and sensitivity degradation from \citet{Breeveld_etal_2011}.  This is done as part of the Swift Optical/Ultraviolet Supernova Archive (SOUSA; \citealp{Brown_etal_2014_SOUSA}).  The updated photometry is available from the Swift SN website\footnote{http://swift.gsfc.nasa.gov/docs/swift/sne/swift\_sn.html}. 
We limit the sample to normal SNe (i.e. no SN~1991T-like, 1991bg-like, or 2002cx-like SNe) with $ 1.0 < $\mb$ < 1.4 $ observed with Swift at the time of the B band maximum light.  This cut is done for two reasons.  First, the models under consideration were based on normal SNe and thus are not as applicable for the other subtypes.  We do not want color differences that might correlate with \mb~to be confused with the color differences from metallicity.  Second, our metallicity concerns focus on normal SNe that are useful for cosmology.
We do not correct for extinction, but will use color-color plots and reddening vectors to show the effect it will have.
% {\bf add a table of the SN properties used in the figures and deltaM15(B) ? }

 For comparison with the observations, spectrophotometry on the HST spectra and model spectra use the revised effective area curves and zeropoints of \citet{Breeveld_etal_2011}.   The wavelength range of the filters is shown in the top panel of Figure \ref{fig_filters}.  The spectrophotometry naturally includes the red tails of the uvw2 and uvw1 filters which transmit a significant amount of optical light when observing very red sources.  In \citet{Brown_etal_2010} an approximate correction (equivalent to an s-correction to an idealized filter with the tails truncated) was utilized to study the dispersion of absolute magnitudes in the UV as might be observed at higher redshift with an optical filter with a sharper transmission cutoff.  
Such corrections can be highly uncertain because they require a spectrum to be assumed or modeled in order to determine the relative fraction coming from the UV and optical portions of the spectrum.  
In comparing with model spectra, such corrections are unnecessary as spectrophotometry will include the effects of the optical tails.  Differences that exist solely in the UV may be diluted by the optical flux, but we show below that the uvw2 and uvw1 still provide unique information.  
In the following we will refer to wavelengths between 2500-4000 \AA~(including the UVOT u and uvw1 filters) as the near-UV (NUV) and 1600-2600 \AA~(including the uvw2 and uvm2 filters) as the mid-UV (MUV). UVOT's optical filters are b and v (4000-6000 \AA).  For our broad color comparisons we will use uvm2 to represent the MUV (because it has very low sensitivity to optical photons), uvw1 to represent the NUV (due to several bright SNe Ia being at or above the saturation limit in Swift's u band), and v-band to represent the optical (for the broadest wavelength coverage).  Other filter choices would give the same general conclusions, and the wavelengths of certain features would make certain filter combinations particularly advantageous.

\begin{figure} 
\vspace{-2cm}
\resizebox{8.0cm}{!}{\includegraphics*{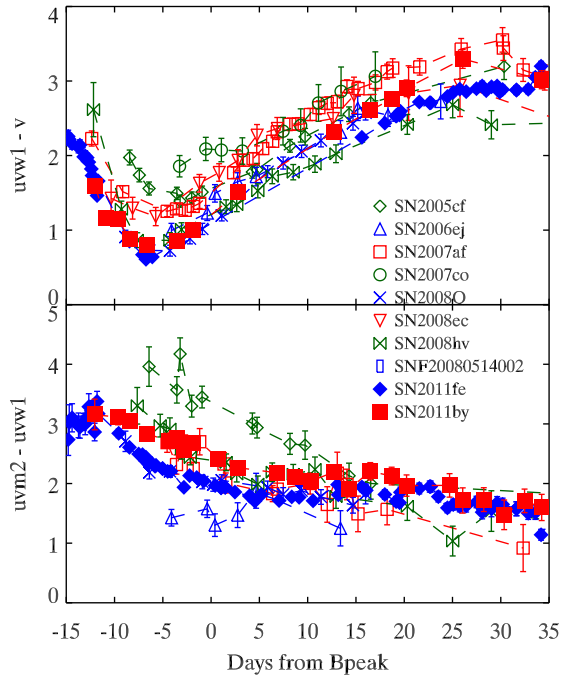}   }
\caption[SN2011by and SN2011fe Comparisons]
        {Color curves of SNe 2011by and 2011fe along with other young, normal SNe Ia \citep{Brown_etal_2014}.  The uvw1-v colors of SNe 2011by and 2011fe are very similar.  The uvm2-uvw1 colors of SNe 2011by and 2011fe are distinct (but small) at all times, but the difference becomes smaller with time.
 } \label{fig_febycolors}    
\end{figure} 

In Figure \ref{fig_febycolors} we show the Swift/UVOT colors of a representative sample of SNe.  
To facilitate comparisons with \citet{Foley_Kirshner_2013},  we highlight the colors of SNe 2011by and 2011fe, which show the same characteristics reported from the spectra: 
The optical and NUV light curves and colors are very similar, differing only in the epochs more than five days before the B band peak.  In the MUV, SN2011fe is consistently bluer.  This difference is qualitatively similar to the colors and absolute magnitudes studied by \citet{Milne_etal_2010} and \citet{Brown_etal_2010}, namely, a low dispersion in the optical and NUV, and an increased dispersion in MUV uvm2 and uvw2.  The growing Swift SN sample shows there could also be a significant variation or bimodal distribution in the NUV-optical colors \citep{Thomas_etal_2011,Milne_etal_2013}.  In this paradigm, SNe 2011by and 2011fe are both classified as NUV-blue SNe \citep{Milne_etal_2013}.  Compared to the larger sample of SNe, SNe 2011by and 2011fe are quite similar in the MUV as well, as the MUV differences between SNe 2011by and 2011fe are small compared to the large scatter seen between other objects.  It is the cause of these similarities and differences which we wish to explore with various theoretical models.

\section{Theoretical Models \label{models}} 

As described in Section 1, many different groups have examined the differences which metallicity and other parameters have on SN observables.  Here we describe in more detail five sets of models for which we have spectra for our comparisons: \citet{Lentz_etal_2000}, \citet{Sauer_etal_2008}, \citet{Walker_etal_2012}, \citet{Blondin_etal_2013}, and \citet{Kromer_Sim_2009}.  We will refer to these hereafter as L00, S08, W12, B13, and K09, respectively.

%%%%%%%%%%%%%%  L00
The L00 models begin with a W7 deflagration model \citep{Nomoto_etal_1984,Thielemann_etal_1986}.  The radiative transfer calculations were performed with the PHOENIX code \citep{Hauschildt_Baron_1999, Hauschildt_etal_1996}.  The models are given at epochs 7, 10, 15, 20, and 35 days after explosion, with luminosity parameters modified to fit optical (not UV) spectra of SN~1994D at -12, -9, -4, 0, and 15 days after B band maximum light.  The metallicity is changed by scaling the number abundance of elements heavier than oxygen in the outer, unburned C+O layer (velocities 14,800-30,000 km s$^{-1}$) by a factor $\zeta$ between 1/30 and 10.  They then renormalize the mass fractions in each layer.  To simulate the nucleosynthetic effect of metallicity changes, they also scale the amount of $^{54}$Fe in the incomplete burning zone (8800-14,800 km s$^{-1}$) in the same way.  Considering the effects separately, the element abundances in the unburned layer have a much stronger effect on the observed spectra.  
Here we use the model spectra where both of these effects are combined, parameterized by the scaling factor $\zeta$. 
%$\epsilon$.  

\begin{figure*} 
\plottwo{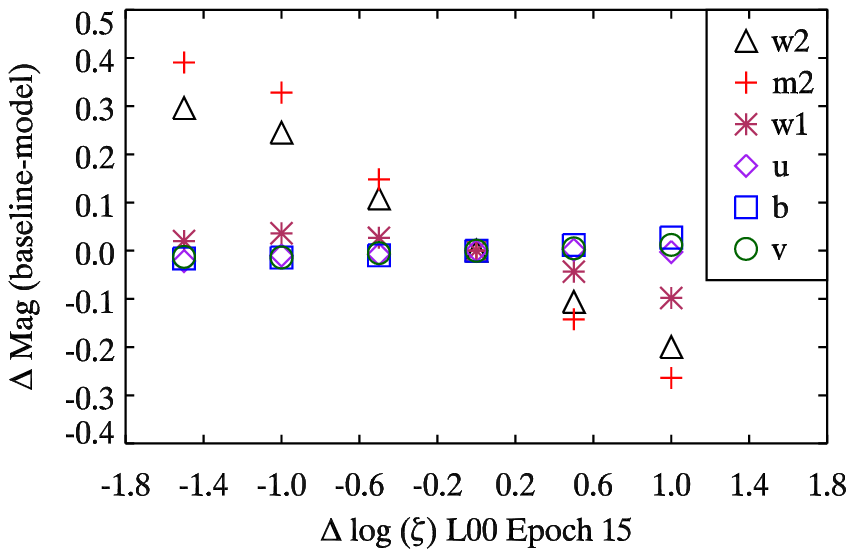} {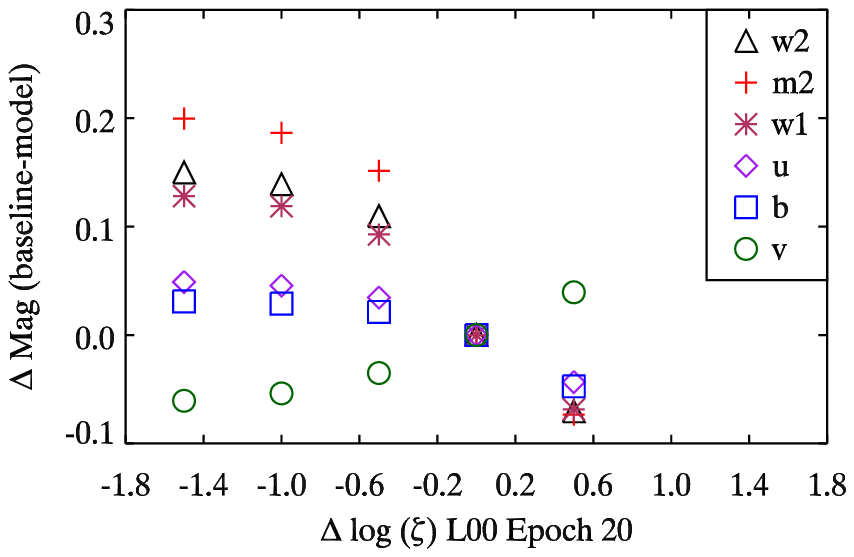} 
\caption[Magnitude differences for Lentz models]
        {Left: Magnitude differences in the UVOT filters for the Lentz models at 15 days after explosion.  These are spectrophotometric magnitudes computed from the model spectra subtracted from the baseline measurements to highlight the differences.  Positive values indicate an increase in flux in that filter relative to the baseline model. 
Right: Magnitude differences in the UVOT filters for the Lentz models at 20 days after explosion.  While the largest differences are still in the MUV filters, there is also significant change in the other filters. 
 } \label{fig_lentzmodelmags}    
\end{figure*}

Figure \ref{fig_lentzmodelmags} shows the spectrophotometry of models corresponding to 15 and 20 days after explosion.  
These and the other model plots show the model photometry subtracted from the baseline model photometry to quantify the differences.  In the day 15 models the MUV brightness increases as $\zeta$ decreases (corresponding to a lower heavy element abundance in the outer layer).  The changes to the optical and NUV are negligible, consistent with the flux ratios shown by \citet{Foley_Kirshner_2013}.  In the right panel, the day 20 models (corresponding to the maximum light epoch of SN~1994D) the changes are more modest and affect all of the filters but with the effect strongest at shorter wavelengths.

\begin{figure*} 
\plottwo{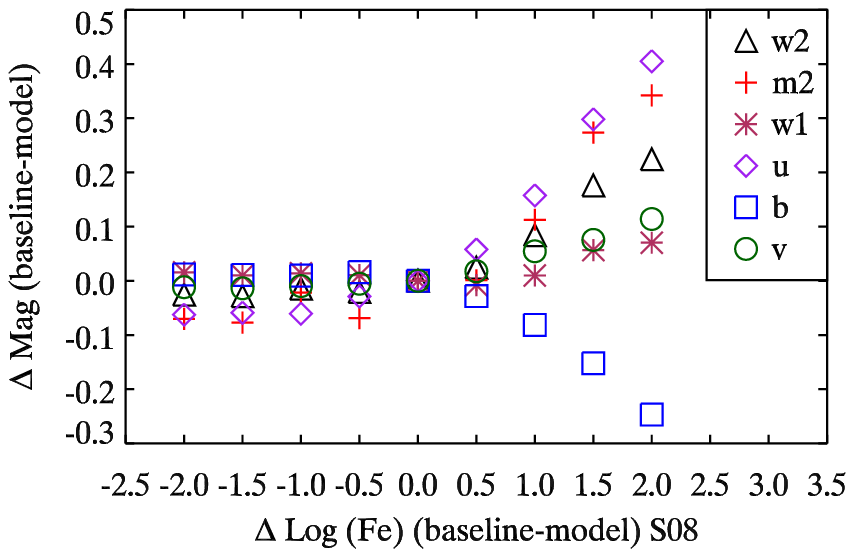}{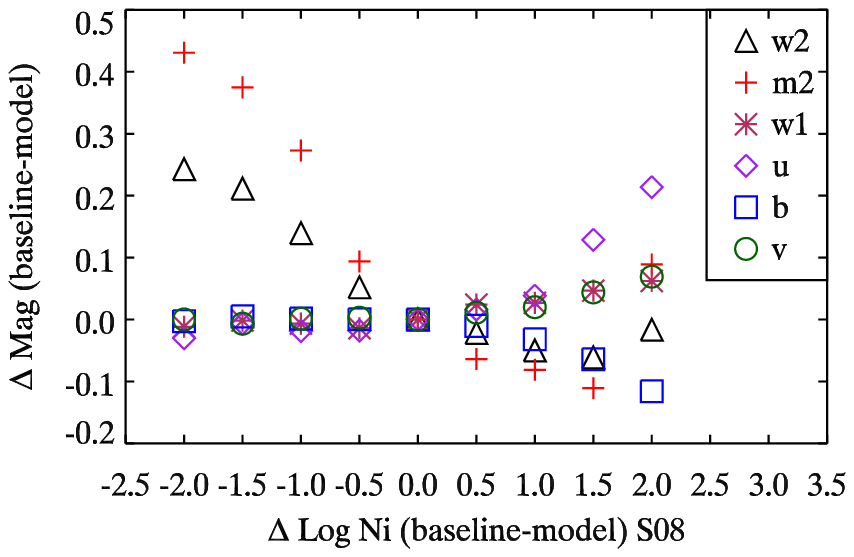}  
\caption[Sauer Metallicity Model Magnitudes]
        {Magnitude differences for different forms of metallicity variations. These are spectrophotometric magnitudes computed from the model spectra subtracted from the baseline measurements to highlight the differences.  The left panel shows the differences varying the amount of Fe.  The right panel shows the differences from varying $^{56}$Ni abundances.  In both cases the UV is much more strongly affected than the optical, but the filters are affected by different amounts for the different variations.
 } \label{fig_sauermetallicitymags}    
\end{figure*}

%%%%%%%%%%%%%%   S08
S08 used the density structure W7 deflagration model \citep{Nomoto_etal_1984,Thielemann_etal_1986} and the Monte-Carlo radiative transfer code of \citet{Mazzali_Lucy_1993}.  The composition and luminosity were tuned so that the emergent spectra matched the UV/optical spectra of SNe 2001eh and 2001ep at 9 days after B-peak (the epoch of the HST UV spectra).  Here we use the models of the moderately declining (\mb=1.41) SN~2001ep.  
The effect of metallicity is probed by changing the abundance of certain elements in the outermost zone with velocities above 14,500 km s$^{-1}$ (extending to 70,000 km s$^{-1}$) at the expense of oxygen.  
They modify the stable Fe, $^{56}$Ni, and Ti and Cr together.  Of these, $^{56}$Ni has the strongest effect in the MUV, though each has a unique signature in the spectral differences which also shows up in the broadband photometry.  
%Figure \ref{fig_sauermetallicitymags} shows the relative spectrophotometry of those models.  
%In the left panel, decreasing the amount of Fe has little effect at any wavelength, 
%but the u and uvm2 bands are brighter (and the b band fainter) for larger amounts of Fe.  
%In the middle panel, the MUV filters are brighter for lower amounts of Ni.  
%As the amount of Ni is increased, the u band is brighter while the MUV and b bands are fainter.  
%For low amounts of Ti and Cr the magnitudes are relatively unchanged while for high values the uvm2 filter is brighter and the uvw1 filter is fainter.  
Figure \ref{fig_sauermetallicitymags} shows the relative spectrophotometry for the Fe and $^{56}$Ni models.  
In the left panel, decreasing the amount of Fe has little effect at any wavelength, 
but the u and uvm2 bands are brighter (and the b band fainter) for larger amounts of Fe.  This increase in UV flux is attributed to a larger number of saturated Fe lines encouraging reverse fluorescence.
In the right panel, the MUV filters are brighter for lower amounts of Ni.  
As the amount of Ni is increased, the u band is brighter while the MUV and b bands are fainter.  
The fact that filters with significant overlap can move in different directions shows that some of the effects are restricted to narrow bands in wavelength.  

\begin{figure*} 
\plottwo{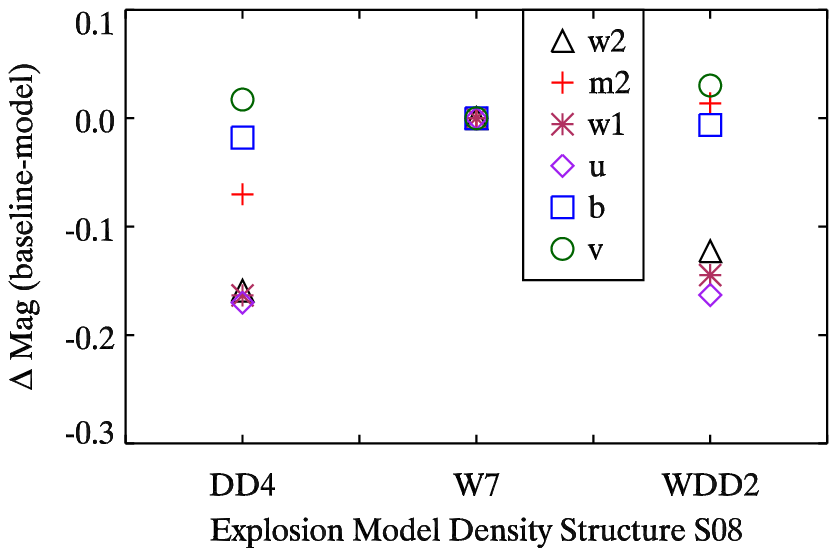}{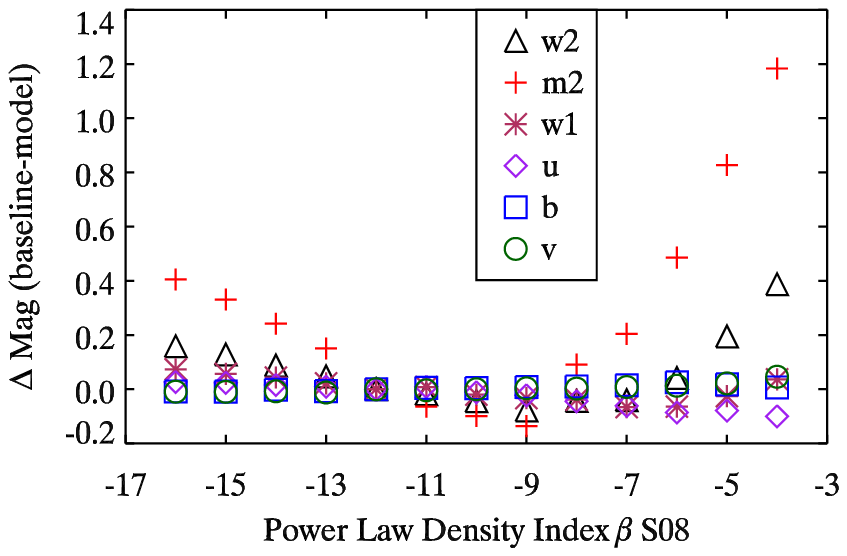} 
\caption[Magnitude differences of for the S08 models]
        {Left: Magnitude differences in the UVOT filters for the S08 models using the density structure from different explosion models.   These are spectrophotometric magnitudes computed from the model spectra subtracted from the baseline measurements to highlight the differences. 
%{\bf add the -12 power law model in this panel as well to see the difference?  }

Right: Magnitude differences in the UVOT filters for the S08 models with the density structure of the outer layers parameterized by a power-law with a variable index.  To isolate the effect of changing the slope of the outer density structure, the $\beta$=-12 model is used as the baseline for this comparison.  Only the mid-UV (uvw2 and uvm2) filters are affected.
 } \label{fig_sauermodelmags}    
\end{figure*}

S08 also compute spectra for different density structures resulting from different explosion models (W7, WDD2, and DD4) and with the density of the outer layers modeled as a power-law with a varying index.  The relative spectrophotometry is shown in Figure \ref{fig_sauermodelmags}.  The NUV flux is significantly weaker for the DD4 and WDD2 models compared to W7.  For varying density profiles, the NUV flux is strongly enhanced for shallower profiles but also enhanced for steeper profiles.  
Clearly the density profile has a strong but non-linear effect on the MUV luminosity.
The wavelength where the differences show up -- the explosion models affecting the NUV and the varying power law in the outer density affecting the MUV -- might prove an important diagnostic in constraining the true density structure.  \citet{Mazzali_etal_2014} had to modify the standard explosion model density structure to fit the UV spectra of SN~2011fe.

%Figure \ref{fig_modelmags} shows the differential magnitudes with respect to the baseline models.  Each set of models has a signature in how the different filter magnitudes behave.  In all cases the optical magnitudes show the smallest change while the MUV filters (uvw2, uvm2) show the greatest change.

%%%%%%%%  W12
The W12 models use the density profile of the W7 deflagration model and delayed detonation models WDD1 and WDD3 \citep{Iwamoto_etal_1999} depending on the luminosity/Ni mass produced.  The radiative transfer is done with a Monte-Carlo code \citep{Mazzali_Lucy_1993}. 
Using an abundance tomography approach \citep{Stehle_etal_2005} to model the UV/optical observations of SN~2005cf, many premaximum epochs are utilized to constrain abundances in the outer layers to create the near-maximum model matched to the observations 0.9 days before B-band maximum.  The metallicity is changed by scaling the abundances of all elements with atomic number above calcium by a factor $\eta$ ranging between 0.05 and 5.  The mass fractions are similarly scaled at the expense of unburnt C/O.  This is done in the outer three layers of the model, corresponding to velocities greater than 13,100 km s$^{-1}$  for the luminosity log($L_{bol}/L_\sun$)=9.6 model.  Plots showing the effect from changing single elements are also given in \citet{Walker_etal_2012}.  They also create models with varying luminosity corresponding to different energies and $^{56}$Ni production.

\begin{figure*} 
\plottwo{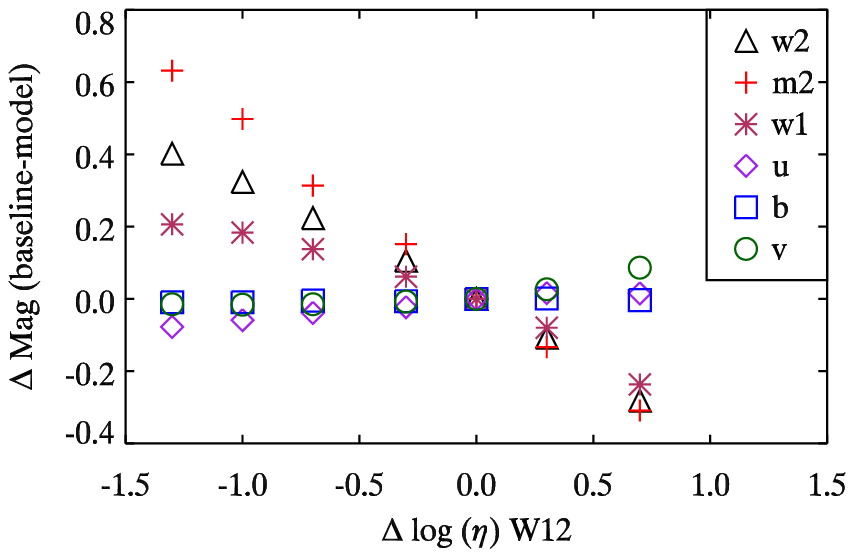} {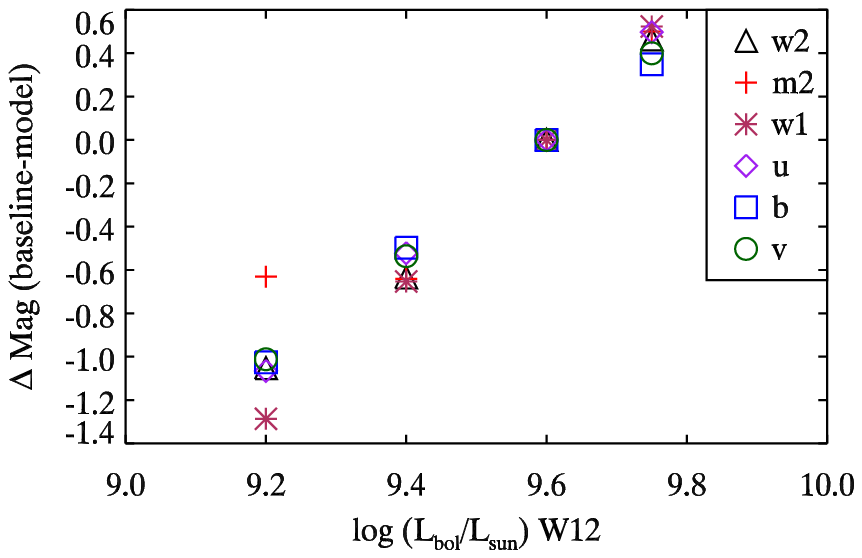} 
\caption[Magnitude differences of for the Walker models]
        {Left: Magnitude differences in the UVOT filters for the Walker models (with fixed log(L)=9.6) with varying metallicity.  These are spectrophotometric magnitudes computed from the model spectra subtracted from the baseline measurements to highlight the differences.  

Right: Magnitude differences in the UVOT filters for the Walker models with varying luminosity and a fixed metallicity.  } \label{fig_walkermodelmags}    
\end{figure*}

The left panel of Figure \ref{fig_walkermodelmags} shows the effect of changing the metal abundance in the outer layers.  For lower metal abundance the MUV filters and the NUV uvw1 filter are brighter, while increasing the metal abundance lowers the flux in those filters.  The right panel shows the different luminosity models. This is clearly reflected in the magnitude differences of about 1.6 mags for most filters.  The lowest luminosity model also shows a significant chromatic difference with the uvw1 relatively fainter and the uvm2 relatively brighter.  Note how the effects cancel each other out leaving the uvw2 magnitude unchanged.

%%%%%%%%%%  B13

\begin{figure} 
\plotone{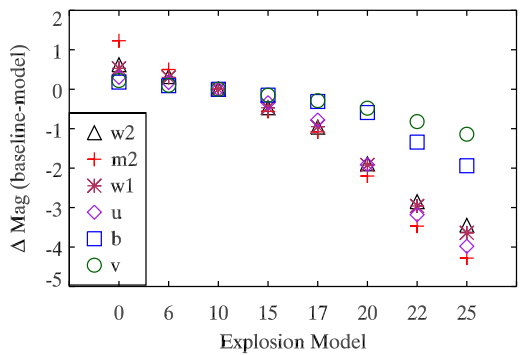} 
\caption[Blondin Model Magnitudes]
        {Magnitude differences (compared to the DDC10 model) for different 1D delayed-detonation models \citep{Blondin_etal_2013}.
See \citet{Blondin_etal_2013} for details about the numbered models.
 } \label{fig_blondinmodelmags}    
\end{figure}

B13 uses a forward modeling approach with a grid of one-dimensional Chandrasekhar-mass delayed detonation models as in \citet{Khokhlov_1991}.  The density at which the transition from deflagration to detonation is varied, resulting in differing amounts of $^{56}$Ni.  The radiative transfer was done using the CMFGEN code \citep{Hillier_Miller_1998,Hillier_Dessart_2012}.  From a larger grid, eight models were published in B13 which matched observed spectra.  Compositions are not varied, but we use these models to see how well a modern forward modeling approach matches the UV photometry of observed SNe.  We also explore the variation of UV colors on the properties of the models.  Figure \ref{fig_blondinmodelmags} shows the model magnitude differences with respect to the DDC10 model which was found to be a good match to SN~2005cf (the same SN on which the W12 baseline model was based).

\begin{figure} 
\plotone{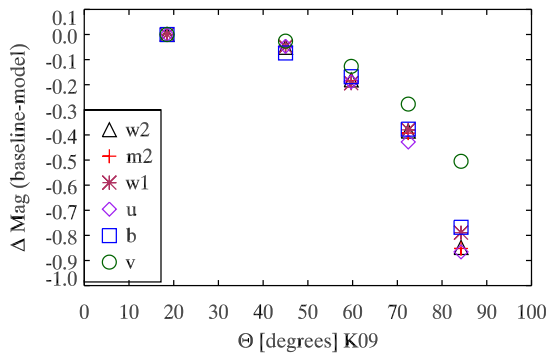} 
\caption[Kromer Model Magnitudes]
        {Magnitude differences  for different viewing angles for an ellipsoidal toy model \citep{Kromer_Sim_2009}. 
 }\label{fig_kromermodelmags}    
\end{figure}

%%%%%%%%%  K09
K09 use an ellipsoidal toy model to explore how the viewing angle of an asymmetric explosion will change the observed properties.  K09 stresses the simplicity and extreme asphericity of the model.  As in K09, we are also less interested in the magnitude of the effect but the relative strength it has on the different filters.  We use the spectral output at maximum light.  Figure \ref{fig_kromermodelmags} shows the differences in magnitude as a function of viewing angle, such that $0^{\circ}$ corresponds to viewing down the semi-major axis and $90^{\circ}$ views down on the equator.

The effect on the magnitudes is less chromatic than for the other models considered.  There is an overall luminosity variation, related to the observed cross section (K09).  The spectrum blueward of 4000 \AA~ is more strongly affected, so the v band is affected less than the others.  Thus B-V or NUV-V colors are affected, but MUV-NUV much less so.  

Before comparing to observations we emphasize that L00 and B13 took a forward modeling approach while the S08 and W12 models were modified to individual SNe with UV observations.  They do not represent all SNe Ia, but we test how well changing the models relative to them reproduces the variation observed in SNe.  We also note that the models considered in this paper do not represent a full sample of the models considered in the literature.  They are not even a full sample of the models considered in the above five studies.  The models used represent the most applicable models available to us.  To allow for comparison, it might be profitable to establish an online database of models comparable to the observational databases currently available via SUSPECT\footnote{http://bruford.nhn.ou.edu/~suspect/index1.html}, WISEREP\footnote{http://www.weizmann.ac.il/astrophysics/wiserep/} \citep{Yaron_Gal-Yam_2012}, and the CfA\footnote{http://www.cfa.harvard.edu/supernova/SNarchive.html} and Berkeley/Filippenko Group SN archives \footnote{http://hercules.berkeley.edu/database/index\_public.html} \citep{Silverman_etal_2012_I}.  The useful outputs to share might range from the density structures used as input to the output spectra and light curves.  To allow theoretical models to be compared with our data, we make our SOUSA photometry available on the Swift SN home page\footnote{http://swift.gsfc.nasa.gov/docs/swift/sne/swift\_sn.html} \citep{Brown_etal_2014_SOUSA}.

\section{Results\label{results}} 

\subsection{Colors \label{colors}}

% made by comparemodels.pro
%\clearpage 
\begin{figure} 
\resizebox{8.8cm}{!}{\includegraphics*{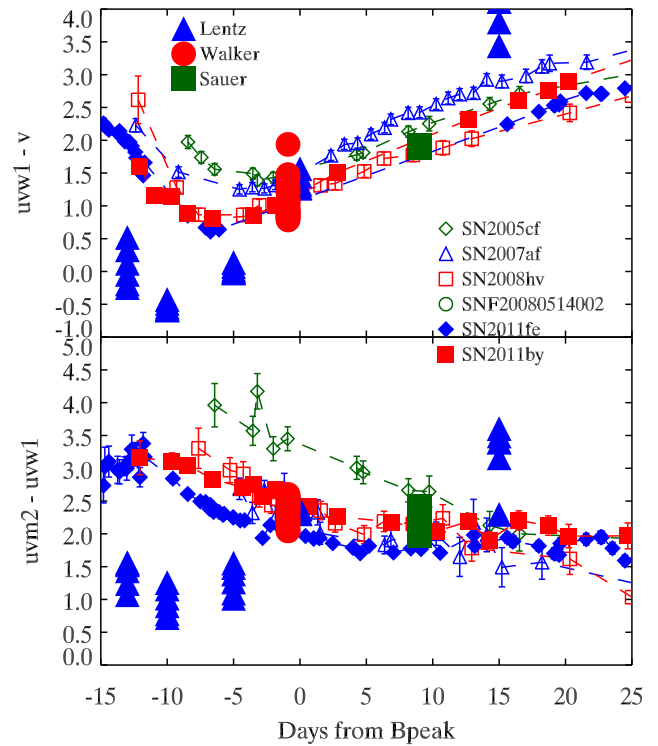}   }
\caption[Results]
        {Color evolution as observed for Swift/UVOT SNe and as predicted by various models.  The color evolution of the Lentz models is faster, stronger and offset from that observed.  The colors of the Walker and Sauer models match most of the SNe (having been tuned to actual observations) but with a scatter inconsistent with the observations. The dispersion in colors and different epochs should constrain the variation of different model parameters.  
 } \label{fig_modelcolorevolution}    
\end{figure}

The models of \citet{Lentz_etal_2000} are at epochs of 7, 10, 15, 20 and 35 days after explosion.  The model spectrum at 20 days was matched to a maximum light spectrum of SN~1994D, so we use that to shift the epochs from days after explosion to days from maximum light.  
%The typical rise time for SNe Ia is 17.4 days, with a spread in times from 13 to 23 days.  
 The \citet{Walker_etal_2012} and \citet{Sauer_etal_2008} models were run for epochs of -1 and 9 days after B maximum, respectively, based on the epochs of the spectra available to them.  We compare the models to the Swift/UVOT observations described above in uvm2-uvw1 and uvw1-v. 
Figure \ref{fig_modelcolorevolution} plots the color evolution of the Swift SNe Ia to the varying-metallicity models L00, S08, and W12.  

\begin{figure} 
\resizebox{8.8cm}{!}{\includegraphics*{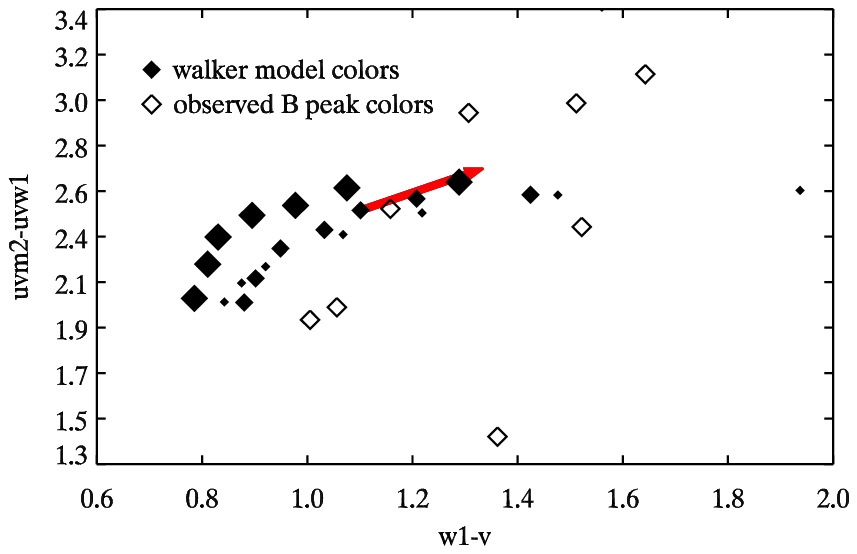}   }
\caption[Results]
        {  Color-color plot of the models and observations.  The colors of the \citet{Walker_etal_2012} models are shown with solid symbols while the Swift/UVOT observed colors at peak are plotted as open symbols.  An arrow shows the reddening vector for the Milky Way \citep{Cardelli_etal_1989} extinction law with E(B-V)=0.1.  
The observed colors show a larger scatter than the models but not in a manner consistent with reddening.
 } \label{fig_walkercolors}    
\end{figure}

The models of \citet{Lentz_etal_2000} are about one magnitude too blue in both colors and evolve much faster, but the shape of the uvw1-v color evolution is qualitatively similar to the observations.  We emphasize that these models were matched to optical spectra and then modified to explore the differences in metallicity.  The \citet{Lentz_etal_2000} W7 models were not  based on UV spectra and are much bluer than observations (e.g. SN~1992A; \citealp{Kirshner_etal_1993}) or more recent models (compare Figures 1, 2, and 3 in \citealp{Baron_etal_2006}).  Similar models based on a delayed detonation scenario are being investigated (M.~Jenks, et al., in preparation).
The \citet{Walker_etal_2012} and \citet{Sauer_etal_2008} models were based on observed UV spectra and better match the observed colors.  For this study we are less interested in the actual colors but rather the scatter in the observed colors and whether it could be accounted for by the variation in model colors.  The L00 and S08 models underestimate the uvw1-v scatter.  The spread of the W12 models is similar to the spread of the observed colors.  This is consistent with the finding of \citet{Maguire_etal_2012} that the scatter in their NUV spectra was consistent with metallicity variations from W12. The W12 models do not cover the spread in the uvm2-uvw1 colors.  The other models also underpredict the scatter in the uvm2-uvw1 colors, though the S08 models are quite comparable at 9 days after max where the observed uvm2-uvw1 colors are more uniform.  Some additional scatter could be the result of extinction, though we see SNe which are bluer than most of the models.  Color-color vectors could differentiate the effects.  As shown in Figure \ref{fig_walkercolors}, the metallicity differences redden the colors in a manner generally similar to dust reddening, but bluer observed colors suggest intrinsic differences beyond that explored by the models.

%\clearpage 

\subsection{Flux Ratios \label{fluxratios}}

\begin{figure*} 
\plottwo{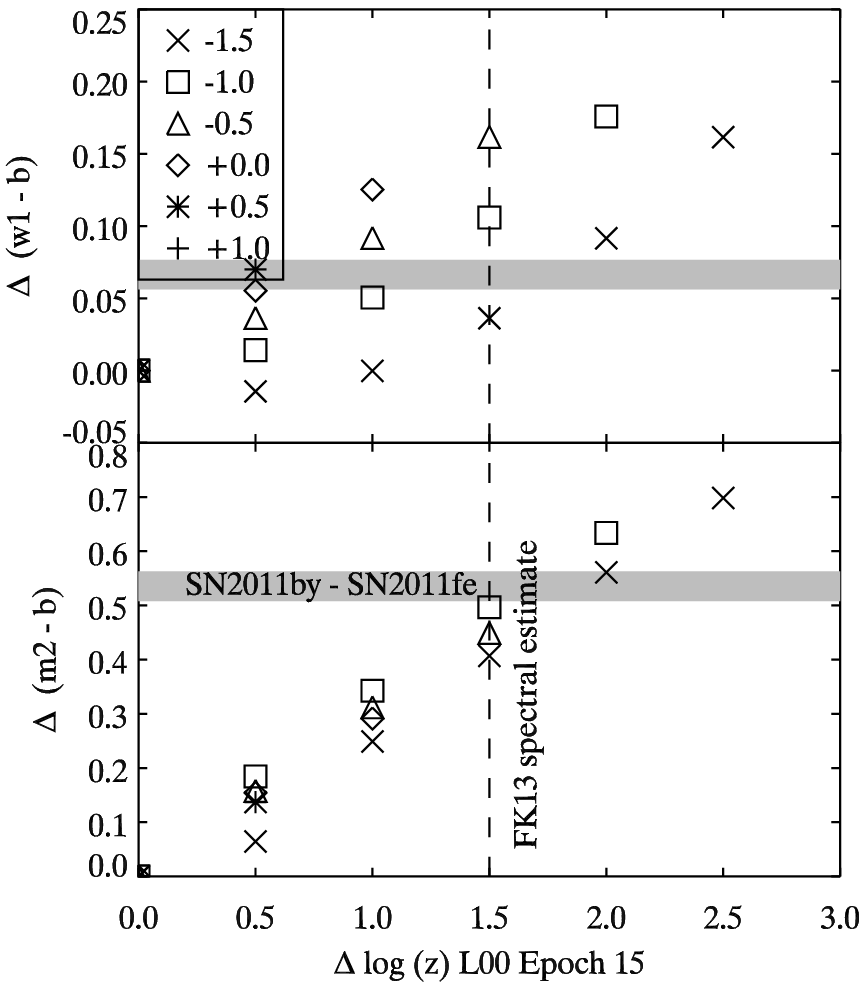}{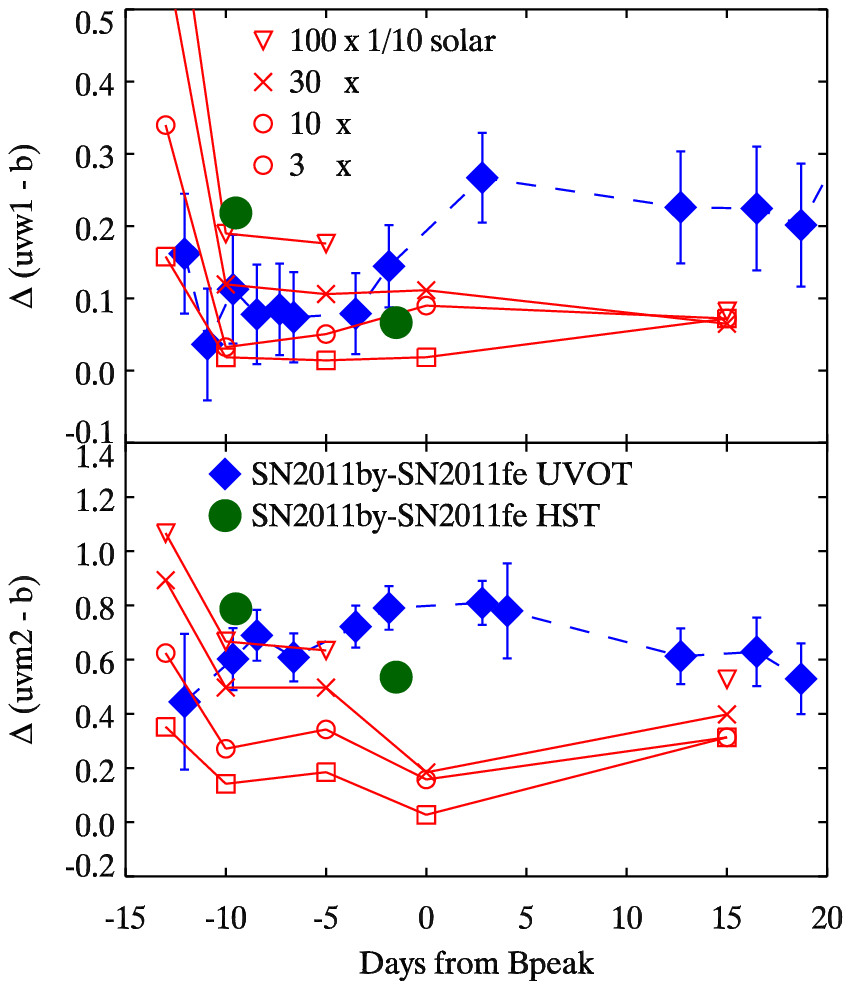}  
\caption[Results]
        {Left Panel: Color differences between the L00 day 15 models versus the metallicity difference between the models.  Each of the models is compared to those of higher (as well as itself, resulting in the points at 0,0).  The baseline (lower metallicity) models used in the comparisons are differentiated by symbols, though the individual points are not important.  What is of interest is the effect on the relative colors resulting from a difference in the metallicity.  The vertical dashed line represents the metallicity difference between SNe 2011by and 2011fe of $\sim$ 30 inferred by \citet{Foley_Kirshner_2013} from HST UV spectra.  The horizontal line represents the color difference of the SNe at peak based on spectrophotometry.  The intersection of the models and uvm2-uvw1 colors show that a comparable metallicity differential could be obtained from uvm2-b photometry alone.  For extreme metallicity differences, the absolute metallicity could also be determined with uvw1-b color errors smaller than 0.05 mag.
Right Panel:  The temporal evolution of the color differences are compared to the UVOT and HST spectrophotometric color differences between SNe 2011by and 2011fe.  The relative differences of the models are consistent before maximum light.  At and after maximum light, however, the observed color differences are not consistent with the \citet{Lentz_etal_2000} models. 
 }
 \label{fig_fluxratiocolors}    
\end{figure*}

To see the difference between two SNe or amongst a group of SNe, one can divide the flux of one by the flux of the other.  This can be done with models as well.  \citet{Foley_Kirshner_2013} used this flux ratio method to attribute the difference between SNe 2011by and 2011fe to a metallicity difference of 1/30 based on the models of \citet{Lentz_etal_2000}.  While some groups have examined the effects on individual line strengths and locations \citep{Walker_etal_2012}, most of the comparisons with observations have relied on broader spectral shapes \citep{Maguire_etal_2012, Foley_Kirshner_2013}.  Such broad spectral shapes can also be measured using photometry.  For one such implementation, we use similar colors as above comparing the MUV to NUV and NUV to optical: uvm2-uvw1 and uvw1-b.  The b band is used here because the HST spectra of SN 2011by do not fully cover the v band.  The color difference between any two models separates out the color effects as a function of the model parameter changed.  The left panel Figure \ref{fig_fluxratiocolors} shows the color differences for the day 15 L00 models (each subtracted from lower metallicity) as a function of the metallicity ratio between the two.  As reported by \citet{Foley_Kirshner_2013}, the flux ratios in the mid-UV are similar between models with the same metallicity ratios regardless of the absolute metallicity.  
The dotted vertical line shows the metallicity difference inferred by \citet{Foley_Kirshner_2013} from dividing the SN~2011fe spectrum by the SN~2011by spectrum and comparing with the spectral flux ratios of the Lentz models.  

The horizontal line shows the color difference of SNe 2011fe and 2011by from HST just before maximum light (the same spectra used by \citealp{Foley_Kirshner_2013}).  The uvw1-b colors of the models contain too much scatter without knowing the absolute metallicity of one of the objects and are consistent with a large spread in metallicity difference from 3-50.  The metallicity difference inferred from the photometry would be a factor of 100, and uncertainties of a few percent in the colors would also allow the FK13 estimate of 30.

The fact that the two lines and the model photometry differences intersect at the same place suggests that UV photometry could put the same constraints on the metallicity differences between SNe as the rarer and harder to obtain UV spectra.  This result assumes that the model differences are the same as those in the SNe.
Such a comparison might also require further optical photometry and spectra to identify SN pairs which are otherwise similar.  The photometric comparison does miss out on small scale differences due to varying line or continuum flux, but so did the spectral comparison of \citet{Foley_Kirshner_2013}.  Studying small scale differences in UV spectra might better reveal the origin of the differences.  The UV photometric differences could then constrain the magnitude of the differences.

\citet{Graham_etal_2015} found that a comparison of the premaximum spectra of SNe 2011by and 2011fe were also consistent with a 1/30 difference in the metallicity.  With the larger temporal coverage of the UVOT photometry of SNe 2011by and 2011fe we can compare the post-maximum behavior.  Similar to the left panel, the right panel of Figure \ref{fig_fluxratiocolors}  shows relative color differences between the models and the observations.  The day 20 spectral models of \citet{Lentz_etal_2000} which were modeled to a maximum light spectrum of SN~1994D show a much depressed variation in the mid-UV.  The post-maximum behavior is also very different than observed.  
While this photometric method appears promising, the \citet{Lentz_etal_2000} models do not accurately reflect the colors in an absolute or relative sense. 
% \citet{Foley_Kirshner_2013} used the Lentz model to infer a metallicity difference.  Combined with xxx they also inferred a difference in the absolute magnitudes seemingly consistent with the observed difference between SNe 2011by and 2011fe.  

% Comparing to a grid of metal content and density structure variations could show which factor dominates the differences between these otherwise quite similar SNe.

%%%%%%%%%%%%%%%%%%%%%%%%%%%%%%%%%%%%%%%%%%%%%%%%%%%% 

%%%%%%%%%%%%
%%%%%%%%%%%%%%%%%%%%%%%%%%%%%%%%%%%%%%%%%%%%%%%%%%%%%%%%%%%%%%%%%%%%%%
\section{Summary\label{discussion}} 

In this paper we have shown the photometric effect of changing the metallicity, density structure, and asymmetry of type Ia SN models.  Metallicity and density structure changes cause photometric differences which generally increase at shorter wavelengths, while asymmetry  changes affect the B-V color and are flat blueward of B.  Compared to Swift SNe, the predicted scatter from metallicity variations is smaller than what is observed, particularly in the mid-UV.  The older \citet{Lentz_etal_2000} models do not predict the color evolution very well, but the temporal differences in the colors and color dispersion suggest multi-epoch modeling and predictions to be an attractive way to resolve degeneracies in the many parameters which effect the UV flux (see also \citealp{Hoeflich_etal_2013} and \citealp{Sadler_etal_2013}).  

Ignoring the absolute color differences in the \citet{Lentz_etal_2000} models, we explore how color differences between SNe could be used to measure metallicity differences as has been claimed for UV spectra \citep{Foley_Kirshner_2013}.  We find the photometric differences between SNe 2011by and 2011fe give the same metallicity differences as the spectra when using the same pre-maximum light models.   The near-peak models form \citet{Lentz_etal_2000}, however, underpredict the mid-UV differences. The \citet{Lentz_etal_2000} models do not reproduce the color differences from multi-epoch photometric comparisons.  While metallicity may indeed be the difference between these similar SNe, the difference found by \citet{Foley_Kirshner_2013} and \citet{Graham_etal_2015} is not supported by our observations.    
Here we have followed \citet{Foley_Kirshner_2013} in assuming that the color differences between SNe 2011by and 2011fe are solely caused by a single parameter under investigation.  The photometric tests show that a number of different parameters strongly affect the UV light.  The large UV dispersion seen in SN observations is likely the product of multiple effects.  With improved models, however, time-series comparisons (with photometry or spectroscopy) might prove very effective in quantifying physical differences between SNe or their progenitors and showing which differences are observationally degenerate.

%There is a modest difference in the NUV-optical colors which persists past maximum light \citep{Milne_etal_2013}.  There is a large dispersion in the MUV-NUV colors which is strongest at early times \citep{Brown_etal_2014}.  It is clear that many intrinsic differences in SNe Ia have strong signatures in the UV.  
Understanding the temporal changes is an important step in understanding the differences amongst SNe Ia, since different effects will likely have different time scales.  
Historically, detailed modeling of SNe relied on high-quality, single epoch UV spectra.  The superbly sampled UV observations of SN~2011fe \citep{Mazzali_etal_2014} allowed the energetics, density structure, and metallicity to be very well constrained.  Modifying those parameters (similar to \citealp{Lentz_etal_2000}, \citealp{Sauer_etal_2008} and \citealp{Walker_etal_2012}) and following the resultant differences in the UV colors would clarify what could make SN~2011by appear so similar in the optical and yet different in the UV \citep{Foley_Kirshner_2013}.  It would also help us better understand the UV differences for the larger sample of SNe observed photometrically in the UV.  Despite the larger photometric sample, additional UV spectral series are needed in order to map out the variations in observed properties.

%%%%%%%%%%%%%%%%%%%%%%%%%%%%%%%%%%%%%%%%%%%%%%%%%%%%%%%%%%%%%%%%%%%%%%%%%%%%%%%%%

\acknowledgements

This work benefitted greatly from discussions with R. Foley and others 
at the 2013 Mitchell Workshop and E. Lentz at the 2013 Fifty One Ergs 
conference at North Carolina State University.  
We would like to thank D. Sauer, S. Blondin, and M Kromer for sharing their models and expertise to this project.
We are also grateful for helpful suggestions from P. Hoeflich.  
This work was supported at Texas A\&M University by NSF grant 497561-0001, 
NASA ADAP grant NNX13AF35G, and the generous support of the Mitchell Foundation.  
This work made use of public data in the {\it Swift} data
archive and the NASA/IPAC Extragalactic Database (NED), which is
operated by the Jet Propulsion Laboratory, California Institute of
Technology, under contract with NASA.  

%%%%%%%%%%%%%%%%%%%%%%%%%%%%%%%%%%%%%%%%%%%%%%%%%%%%%%%%%%%%%%%%%%%%%%%%%%%%%%%%%

%%%%%%%%%%%%%%%%%%%%%%%%%%%%%%%%%%%%%%%%%%%%%%%%%%%%%%%%%%%%%%%%%%%%%%%%%%%%%%%%%

\bibliographystyle{../apj}

%% the following is the path to you *.bib file
%% (you do not need to enter the ``.bib'' extention)
%../bibtex}

%%%%%%%%%%%%%%%%%%%%%%%%%%%%%%%%%%%%%%%%%%%%%%%%%%%%%%%%%%%%%%%%%%%%%%

%
%%%%%%%%%%%%%%%%%%%%%%%%%%%%%%%%%%%%%%%%%%%%%%%%%%%%%%%%%%%%%%%%%%%%%%

\end{document}